\documentclass[fleqn,usenatbib]{mnras}

\usepackage{newtxtext,newtxmath}

\usepackage[T1]{fontenc}
\usepackage{ae,aecompl}

\usepackage{graphicx}	
\usepackage{amsmath}	
\usepackage{amssymb}	

\title[GW emission \textcolor{black}{in TDEs}]{Gravitational wave emission \textcolor{black}{from unstable accretion discs in tidal disruption events}}

\author[M. Toscani et al.]{
Martina Toscani,$^{1}$\thanks{E-mail: martina.toscani@unimi.it}
Giuseppe Lodato$^{1}$
and Rebecca Nealon$^{2}$
\\
$^{1}$Dipartimento di Fisica, Universit\`a Degli Studi di Milano, Via Celoria, 16, Milano, 20133, Italy\\
$^{2}$Department of Physics and Astronomy, University of Leicester, University Road, Leicester, LE1 7RH, UK\\
}

\date{Accepted XXX. Received YYY; in original form ZZZ}

\pubyear{2019}

\begin{document}
\label{firstpage}
\pagerange{\pageref{firstpage}--\pageref{lastpage}}
\maketitle

\begin{abstract}
Gravitational waves can be emitted by accretion discs if they undergo instabilities that generate a time varying mass quadrupole. In this work we investigate the gravitational signal generated by a thick accretion disc of $1 $M$_\odot$ around a static super-massive black hole of $10^{6}$M$_\odot$, assumed to be formed after the tidal disruption of a solar type star. This torus has been shown to be unstable to a global non-axisymmetric hydrodynamic instability, the Papaloizou-Pringle instability, in the case where it is not already accreting and has a weak magnetic field. We start by deriving analytical estimates of the maximum amplitude of the gravitational wave signal, with the aim to establish its \textcolor{black}{detectability} by the Laser Interferometer Space Antenna (LISA). Then, we compare these estimates with those obtained through a numerical simulation of the torus, made with a 3D smoothed particle hydrodynamics code. Our numerical analysis shows that the \textcolor{black}{measured} strain is two orders of magnitude lower than the maximum value obtained analytically. However, accretion discs affected by the Papaloizou-Pringle instability may still be interesting sources for LISA, if we consider discs generated after deeply penetrating tidal disruptions of main sequence stars of higher mass.
\end{abstract}




\section{Introduction}
Tidal disruption events (TDEs) occur when a star wanders too close to a super-massive black hole (SMBH), getting disrupted by the tidal field of the hole. After the disruption, roughly half of the star circularizes and forms a disc-shaped structure that accretes onto the SMBH, while the other half escapes on hyperbolic orbits with different orbital energies. These events are important electromagnetic sources, in UV-optical (e.g. \citealt{Gezari:08aa}, \citealt{Komossa:08aa}, \citealt{Gezari:09aa}, \citealt{Gezari:17aa}), in X-rays (e.g. \citealt{Bade:96aa}, \citealt{Komossa:99ab}, \citealt{Komossa:99aa}, \citealt{Greiner:00aa}), in radio (e.g. \citealt{Zauderer:11aa}) and also in $\gamma$-rays (e.g. \citealt{Bloom:11aa}, \citealt{Levan:11aa} and \citealt{Cenko:12aa}). The bolometric lightcurve of these events is expected to decrease with time as $t^{-5/3}$, like illustrated by \citet{Rees:88aa} and \citet{Phinney:89aa}. However, recent studies (\citealt{Lodato:09aa}, \citealt{Guillochon:13aa}) have shown that this trend is reached only at late times and that lightcurves in specific bands might show a different evolution (e.g. \citealt{Lodato:11aa}).\\
\indent The mechanism of accretion onto the central object may affect the formation of the accretion disc. In particular there are three parameters that we need to consider in this process: the circularization time, $t_{\rm circ}$, the viscous accretion time $t_{\rm visc}$ and the radiative cooling time $t_{\rm cool}$ (\citealt{Evans:89aa}, \citealt{Bonnerot:16aa}). If $t_{\rm visc} > t_{\rm circ}$, the accretion process starts only when the disc is formed. In this scenario, we have two possibilities: if $t_{\rm cool}<t_{\rm circ}$ the disc is thin, otherwise the disc is thick, so we have a torus (see, e.g., \citealt{Lodato:07aa}).\\
\indent If we consider a thick disc, the evolution of the system may undergo a global non-axisymmetric hydrodynamic instability, that is the Papaloizou-Pringle instability (PPI), first described by \citet{Papaloizou:84aa}. For this instability to arise, the torus needs to have a shallow specific angular momentum profile and a well-defined inner and outer radii. Recently, \citet{Nealon:18aa}, using initial conditions motivated by \citet{Bonnerot:16aa}, have shown that a torus formed after a TDE may be susceptible to the PPI, with a frequency close to the Keplerian one and a radius approximately two times the pericenter of the initial stellar orbit. Additionally, for very low initial magnetic fields\textcolor{black}{,} they also suggested that the PPI may drive super-Eddington accretion onto the SMBH faster than the magneto rotational instability (MRI). \textcolor{black}{Around the same time, magnetohydrodynamic simulations of PPI susceptible tori by \citet{Bugli:18aa} illustrated that the presence of magnetic fields is able to quickly suppress the PPI.}\\
\indent The PPI generates a time dependent density distribution, that leads to the emission of gravitational waves (GWs). These waves might be detected by the Laser Interferometer Space Antenna (LISA)\footnote{\url{https://www.elisascience.org}}.To date, the GW signal from unstable accretion discs formed during TDEs has not been investigated yet. Anyway there are few studies on the GW emission associated with the phase of tidal disruption of the star. Some of these are based on full general relativistic hydrodynamics codes (see, e.g., \citealt{Haas:12aa}, \citealt{Anninos:18aa}), other on smoothed particle hydrodynamics (SPH) codes (\citealt{Rosswog:09aa}, \citealt{Kobayashi_2004}).\\
\indent \citet{Haas:12aa} focus on ultra-close TDEs of white dwarfs (WDs) by a rotating intermediate-massive black hole (IMBH). They assume to have a $10^3\text{M}_{\sun}$ BH, with $1\text{M}_{\sun}$ WD of radius $6000\,\text{km}$ on a parabolic orbit around it, at an average distance from us of $30\,\text{kpc}$. They find that the signal is a burst, with an amplitude $\approx 10^{-18}$ and frequency of few Hz. Moreover, they illustrate that the BH spin does not affect the GW signal in a significant way. \citet{Anninos:18aa} perform simulations of $0.2-0.6\text{M}_{\sun}$ WD on a parabolic orbit around a static $10^3-10^4 \text{M}_{\sun}$ BH, assuming that the source is at an average distance of $10\,\text{Mpc}$. They derive that the GW strain is $\approx 10^{-22}$ and the frequency of the GW signal is $10^{-2}-10^{-1}\,\text{Hz}$, so these sources might be possible interesting targets for LISA. \\
\indent On the other side, \citet{Rosswog:09aa}, with an SPH code, have simulated WDs tidally disrupted by IMBHs obtaining the same conclusions as \citet{Anninos:18aa}. Instead \citet{Kobayashi_2004}, using an SPH code implemented with general relativity, have investigated the GW signal emitted during TDEs of main sequence (MS) and helium stars around a SMBH (both static and rotating) of mass $10^6\text{M}_{\sun}$, showing that it might be detectable by LISA, in the case of strong encounters and assuming that these events take place at a distance $\leq 20 \,\text{Mpc}$. They have also shown that the signal is in most part insensitive to the particular equation of state and structure of the star. Moreover, they have calculated that, for a solar type star, the peak value for the GW strain is around $\sim 10^{-22}$, in the case of static SMBHs, while it is approximately one order of magnitude higher, $\sim 10^{-21}$, in the case of rotating holes. They also have studied the signal analytically, approximating the star with a point-mass particle, and they have found that the numerical results are well \textcolor{black}{described by this approximation}. \textcolor{black}{In the field of GW astronomy, all these works are important since underline the basic features of the GW emission associated with TDEs. In particular they all show that the strain associated with the phase of disruption has a similar shape, independent on the particular structure of the star. In the context of this paper, where we focus on disruption of solar type stars from SMBHs, the most relevant work is that of \citet{Kobayashi_2004}, of which our work can be seen as an extension to describe what happens after the disruption itself}.\\
\indent \textcolor{black}{As for the GW signal generated by the PPI instead}, \citet{Kiuchi:11aa} have illustrated that the PPI in thick self-gravitating tori may produce a detectable signal for GW detectors, both at high frequencies ($100\sim 200\,\text{Hz}$) in the case of stellar BHs, and in the low frequencies range (mHz) in the case of SMBHs. For this latter case, which is the one that LISA might observe, they have in particular considered a system formed by a SMBH of $10^6$M$_{\sun}$ and a thick accretion disc of $\sim 10^5$M$_{\sun}$, system that they assume it might have formed by the collapse of a super-massive star. The orbital radius of the torus is $\approx 10^7 \,\text{km}$ and the system is at a distance of $10 \,\text{Gpc}$. They have made both a numerical study, with a general relativistic grid code, and an analytical study, finding that, if they assume to have the PPI with one over-density, the GW signal peak could reach $10^{-19}-10^{-18}$, with a frequency of  $\sim 10^{-3} \, \text{Hz}$. While \citet{Kobayashi_2004} have found a good \textcolor{black}{agreement} between the numerical and the analytical results for the GW signal from the stellar disruption phase, for the signal generated by an unstable accretion disc \citet{Kiuchi:11aa}'s analytical estimates are one order of magnitude higher than their numerical results.\\
\indent In this paper we link the previous work of \citet{Kobayashi_2004} and \citet{Kiuchi:11aa}, since we investigate the GW signal from tori formed by TDEs \textcolor{black}{of MS stars}. In particular, we start from $1\text{M}_{\sun}$ TDE remnant unstable to the PPI, as shown in \citet{Nealon:18aa}, with no self-gravity, around a non-rotating $10^6\text{M}_{\sun}$ SMBH. We study this signal both through an analytical and a numerical analysis.\\
\indent The structure of the paper is the following. In Section~\ref{sec:2} we illustrate the theory behind our work, i.e. TDE physics, the main features of the PPI and the basics of GW emission. In Section~\ref{sec:3} we describe our analytical estimates, while in Section~\ref{sec:4} the numerical study. In Sections~\ref{sec:5} and~\ref{sec:6} we discuss our results and give our conclusions respectively.

\section{Theory}
\label{sec:2}
\subsection{Tidal disruption physics}
\label{ssec:tde}
Here we consider a standard scenario for stellar disruptions by a SMBH, where a solar type star of mass $M_*$ and radius $R_*$ is on a parabolic orbit around the hole of mass $M_{\rm h}$. Let us suppose to work under the impulse approximation, i.e. the interaction between the star and the SMBH takes place only at the pericenter of the stellar orbit, $r_{\rm p}$. To determine the minimum approach at which the star can be tidally disrupted, we have to equate the gravitational acceleration on the stellar surface due to the self-gravity, $g_{\rm sg}$,
\begin{align}
    g_{\rm sg}=\frac{GM_*}{R_*^2},
    \label{eq:sg_acc}
\end{align}
where $G$ is the gravitational constant, to the tidal field exerted on the star, $g_{\rm t}$,
\begin{align}
    g_{\rm t}\approx \frac{\partial g_{\rm h}}{\partial r}R_*=-\frac{\partial}{\partial r}\left( \frac{GM_{\rm h}}{r^2}\right)R_*=\frac{2GM_{\rm h}}{r^3}R_{*},
    \label{eq:t_acc}
\end{align}
where $g_{\rm h}$ is the gravitational acceleration due to the presence of the central object and $r$ is the distance between the centre of mass of the star and the SMBH. Comparing equation~(\ref{eq:sg_acc}) and (\ref{eq:t_acc}), we obtain that the two fields are comparable at a distance $r_{\rm t}$, called tidal radius, given by
\begin{align}
    r_{\rm t}=R_*\left( \frac{M_{\rm h}}{M_*}\right)^{1/3}\textcolor{black}{\simeq} 7\times 10^{12} r_*\left(\frac{M_6}{m_*} \right)^{1/3}\,\text{cm},
    \label{eq:t_rad}
\end{align}
where we have introduced the dimensionless parameters $m_*$, $r_*$ and $M_6$, such that $M_*=m_*\,\text{M}_{\sun}$, $R_*=r_*\,\text{R}_{\sun}$ and \mbox{$M_{\rm h}=M_6\times 10^6 \,\text{M}_{\sun}$}.\\
\indent Thus, the disruption of the star occurs when \mbox{$r_{\rm p} \leq r_{\rm t}$}. However, for deeply penetrating events, the star may be directly swallowed by the hole. This occurs when
\begin{align}
    r_{\rm p}\leq 2r_{\rm g}\textcolor{black}{=}\frac{2GM_{\rm h}}{c^2}=3M_{6}\times 10^{11}\,\text{cm},
\end{align}
where $r_{\rm g}$ is the gravitational radius of the SMBH and $c$ is the speed of light. To quantify how close the star is with respect to the SMBH and consequently the strength of the disruption, we define a dimensionless parameter, called the penetration factor $\beta$, as
\begin{align}
    \beta=\frac{r_{\rm t}}{r_{\rm p}}.
    \label{eq:pen_fac}
\end{align}
Given the two constraints above, that $r_{\rm p}\leq r_{\rm t}$ and that $r_{\rm p}>2r_{\rm g}$, $\beta$ can vary in the range
\begin{align}
1\leq\beta\textcolor{black}{\lesssim} 20\times r_* M_{6}^{-2/3}m_{*}^{-1/3}.
\label{eq:beta}
\end{align}

\subsection{The Papaloizou Pringle instability}
\label{ssec:ppi}
The PPI was first studied by \citet{Papaloizou:84aa} and \citet{Blaes:86aa}, but a more simplified description of the PPI is covered in \citet{Pringle:07aa}. They consider a cylindrical flow of an incompressible fluid, with no $z$-dependence, that satisfies the Rayleigh criterion. In particular, they assume that this fluid has a rotational velocity, $\Omega$, given by
\begin{align}
  \Omega(R)=\Omega_0\left(\frac{R_0}{R}\right)^{2} ,
\end{align}
where $R$ is the cylindrical radius, $R_0$ is a reference radius and $\Omega_0$ is the angular velocity at $R_0$.
The solution of the perturbation equation for this flow is given by (e.g. \citealt{Blaes:86aa}, \citealt{Pringle:07aa})
\begin{align}
\frac{(\omega+m\Omega_-)^2+mg_-/R_-}{(\omega+m\Omega_+)^2+mg_+/R_+}=\left(\frac{R_+}{R_-} \right)^{2m}\frac{(\omega+m\Omega_-)^2-mg_-/R_-}{(\omega+m\Omega_+)^2-mg_+/R_+},
\label{eq:pert_eq}
\end{align}
where $\omega$ is the mode frequency, $m$ is the azimuthal wavenumber, $R_-$ and $R_+$ are the inner and outer radii respectively and $g$ is the effective gravity
\begin{align}
g(R)=\frac{GM}{R_0^2}\left[ \frac{R_0}{R}\right]^3\left[1-\frac{R}{R_0} \right].
\end{align}
Equation~(\ref{eq:pert_eq}) tells us that the growth rate of the unstable modes is independent of the mass of the torus, instead depending on the geometry of the torus as defined by the inner and outer boundaries.\\
\indent One of the solutions of equation~(\ref{eq:pert_eq}) is an unstable growing mode; this takes place when there is an exchange of angular momentum and energy between a \textcolor{black}{\textit{negative-energy wave}}, that travels from the inner edge of the torus \textcolor{black}{(where $\omega < \Omega$)}, with a wave that travels from the outer edge \textcolor{black}{(where $\omega > \Omega$)}, in the opposite direction \textcolor{black}{with respect to the local medium} but with the same angular phase speed of the \textcolor{black}{\textit{negative-energy wave}}. \textcolor{black}{This redistribution of momentum and energy happens at the corotation radius $R_{\rm cor}$, defined as the radius where $\omega=\Omega$}. When this happens, we have the PPI. In particular \citet{Nealon:18aa} have found that, for the thick torus they have studied, the fastest unstable growing mode corresponds to $m=1$ and manifests as one over\textcolor{black}{-}dense region forming in the torus. 

\subsection{Gravitational wave emission}
\label{ssec:gws}
In general relativity (GR), Einstein's equations, under the weak field approximation, become a wave equation plus a gauge condition\footnote{The Greek indices range from 0 to 3 (the four space-time coordinates), while the Latin indices from 1 to 3 (only spatial coordinates).} \citep{Einstein:18aa}
\begin{align}
    &\Box \bar{h}_{\mu\nu}=-\frac{16\pi G}{c^4}T_{\mu\nu}, \label{eq:ein_wave}\\
    &\partial^{\mu}\bar{h}_{\mu\nu}=0,
\end{align}
where $T_{\mu\nu}$ is the stress-energy tensor and $\bar{h}_{\mu\nu}$ is defined as
\begin{align}
    \bar{h}_{\mu\nu}=h_{\mu\nu}-\frac{1}{2}\eta_{\mu\nu}h,
\end{align}
where $h_{\mu\nu}$ are small perturbations of the Minkowskian metric $\eta_{\mu\nu}$ and $h$ is trace of $h_{\mu\nu}$. If we assume that (e.g. \citealt{Buonanno:07aa}) 
\begin{enumerate}
    \item the internal motion of the source is slow compared to the speed of light,
    \item the self gravity of the source is negligible,
    \item the signal is detected at a distance $D$ very far from the source,
    \item the Transverse-Traceless gauge holds, where only two components of $h_{\mu\nu}$ are independent,\footnote{In the TT gauge we have $\bar h_{ij}$=$h_{ij}$.}
\end{enumerate}
the solution of the wave equation may be written as
\begin{align}
{h}^{\rm TT}_{ij}(t,\pmb x)\simeq\frac{4 G}{Dc^4}\Lambda_{ij,kl}(\pmb n)\int_{|x'|<s} \,\text{d}^{3}x'\,T^{kl}\left(t-\frac{D}{c}+\frac{x'\cdot\pmb n}{c};x' \right),
\label{eq:quad}
\end{align}
where $\Lambda_{ij,kl}$ is the TT-operator, $s$ is the typical size of the system and $\pmb n$ is the direction of propagation of the wave. With the previous assumptions, it is possible to write a multipole expansion of $h^{\rm TT}_{ij}$; however, the monopole term is
zero (due to the conservation of mass of the system), and the dipole term is also zero (due to the
linear momentum conservation). For these reasons, the first non-vanishing term in the expansion is the quadrupole term and so, after some algebraic passages (see, e.g., \citealt{Buonanno:07aa}), equation~(\ref{eq:quad}) reads
   \begin{align}
h^{\rm TT}_{ij}(t,\textbf{n})=\frac{1}{D}\frac{2G}{c^4}\Lambda_{ij,kl}(\textbf{n})\ddot M^{kl}\left( t-\frac{D}{c}\right).
\label{eq:quad2}
\end{align}
where $\ddot M^{kl}$ is the second time derivative of the moment of inertia of the system, $M^{kl}$, defined as
\begin{align}
    M^{kl}=\frac{1}{c^2}\int\,\text{d}^3x \,T_{00} x^{k}x^{l},
    \label{eq:mom_in}
\end{align}
with $T_{00}/c^2=\rho$. \\
\indent As shown in \citet{Buonanno:07aa} and \citet{Maggiore:07aa}, from equation~(\ref{eq:quad}) we can derive the expression for a wave propagating in any direction. In particular, if we consider a wave along the $\pmb z$ direction, we get the following expressions for \textcolor{black}{$h_{11}\doteq h_+$} and \textcolor{black}{$h_{12}\doteq h_\times$} (see \citealt{Maggiore:07aa})
\begin{align}
    h_+&=\frac{G}{Dc^4}(\ddot M_{11}-\ddot M_{22}),\label{eq:h+}\\
    h_{\times}&=\frac{2G}{Dc^4}\ddot M_{12}\label{eq:hx}.
\end{align}
The GW strain can be calculated as
\begin{align}
    h\approx (h_+^2 + h_\times^2)^{1/2}.
    \label{eq:strainpeak}
\end{align}
\indent To make a simple estimate of the GW amplitude associated to a source, we can approximate equation~(\ref{eq:strainpeak}) using the expression given in \citet{Thorne:87aa},
\begin{align}
 h\simeq\frac{1}{D}\frac{4G}{c^2}\frac{E_{\rm kin}}{c^2},
\label{eq:strain}
\end{align}
where $E_{\rm kin}$ is the kinetic energy of the moving source. In order to see if this peak value may be detectable by LISA, we need to compare it with the sensitivity curve of the instrument. This curve (see \citealt{Amaro-Seoane:17aa}) is calculated in terms of the characteristic amplitude of the noise, $h_{\rm n}$, as a function of the frequency, $f$. In particular, $h_{\rm n}(f)$ is defined as (\citealt{Moore:04aa}, \citealt{Maggiore:18aa})
\begin{align}
    h_{\rm n}(f)^2=fS_{\rm n}(f),
    \label{eq:noise_strain}
\end{align}
where $S_{\rm n}(f)$ is the noise spectral density. The quantity related to the GW signal that we compare to $h_{\rm n}(f)$ is the characteristic amplitude of the signal, $h_{\rm c}$, defined as (\citealt{Maggiore:18aa})
\begin{align}
    |h_{\rm c}(f)|^2=4f^2|\tilde{h}(f)|^2,
    \label{eq:car_strain}
\end{align}
where $\tilde{h}(f)$ is the Fourier transform of the strain.

\section{Method}
\label{sec:3}
In this work we study the GW signal associated to $1$M$_{\sun}$ TDE remnant, simulated by \citet{Nealon:18aa}, unstable to the PPI with $m=1$. They assume that the remnant is not magnetised and that the star disrupted by the SMBH is a solar type star, with $\beta=5$, as in \citet{Bonnerot:16aa}. They study the evolution of the system for 20 orbits, although the growth of the unstable PPI modes is expected to be suppressed by the MRI before this.\\
\indent Our work is divided into two parts: an analytical study, to have some first estimates of the expected signal, and a numerical study, to derive the GW strain associated to the torus.

\subsection{Analytical study}
For the analytical estimates, we proceed as follows. The PPI involves a displacement of a mass $\approx 1\text{M}_{\sun}$, moving roughly on a Keplerian orbit around the SMBH at a distance $\approx 2r_{\rm p}$ (see \citealt{Bonnerot:16aa}). A similar argument has been made by \citet{Kobayashi_2004}, when considering the GW signal of the disruption phase. Indeed, the GW signal associated to the stellar disruption can be estimated through equation~(\ref{eq:strain}), expressing $E_{\rm kin}$ as
\begin{align}
    E_{\rm kin}=\frac{1}{2}M_*v^2_{\rm kepl}=\frac{1}{2}M_*\frac{GM_{\rm h}}{r_{\rm p}},
    \label{eq:enkinetic}
\end{align}
where $v_{\rm kepl}$ is the Keplerian velocity of the star. Thus, substituting equation~(\ref{eq:enkinetic}) into equation~(\ref{eq:strain}), we find that
\begin{align}
    h=\frac{r_{\rm s*}r_{\rm g h}}{Dr_{\rm p}},
    \label{eq:h_simpl}
\end{align}
where $r_{\rm s*}$ is the Schwarzschild radius of the star and $r_{\rm g h}$ is the gravitational radius of the hole. We expect the frequency of the signal to be, approximately, the Keplerian frequency 
\begin{align}
    f=\frac{1}{2\pi}\left(\frac{GM_{h}}{r^3_{\rm p}}\right)^{1/2}.
    \label{eq:freq}
\end{align}
In particular, if we consider a MS star, for which it is reasonable to assume that mass and radius scales in the same way, i.e.
\begin{align}
 {\frac{M_*}{M_{\sun}}} \approx {\frac{R_*}{R_{\sun}}} \Rightarrow m_*\approx r_* ,   
 \label{eq:scale_rel}
\end{align}
 we have that equation~\eqref{eq:h_simpl} and equation~\eqref{eq:freq} become 
\begin{align}
 &h\approx 10^{-22}\times\beta\left(\frac{D}{16 \,\text{Mpc}} \right)^{-1}m_{*}^{1/3}M_6^{2/3} \label{strain_ms},\\
 &f\approx 10^{-4}\,\text{Hz}\times \beta^{3/2}m_*^{-1} \label{eq:freq_ms}.
\end{align}
\indent To compare these estimates with the LISA sensitivity curve, we need to derive $h_{\rm c}$ through equation~\eqref{eq:car_strain}. It can be shown (\citealt{Maggiore:18aa}) that this equation may be expressed in a more useful way as 
\begin{align}
    h_{\rm c}\approx h\mathcal{N}^{1/2}_{\rm c},
    \label{eq:strain_cycle}
\end{align}
where $\mathcal{N}_{\rm c}$ is the number of cycles spent in the detector bandwidth $[f_{\rm min}, f_{\rm max}]$ (\citealt{Maggiore:07aa}). Since our source is monochromatic, we can approximate this quantity as 
\begin{align}
    \mathcal{N}_{\rm c}\approx f\tau,
    \label{eq:numbercycle}
\end{align}
where $\tau$ is the emission time (\citealt{Colpi:17aa}).\\
\indent With our analytical estimates we are assuming the \textit{raw strain} of the GWs to be the same in the case of TDEs and unstable accretion discs. However, we see that this is not true for $h_{\rm c}$, since in the case of TDEs (as explored by \citealt{Kobayashi_2004}) we have a burst signal with a duration that is simply the inverse of the frequency. This means that, in this scenario, $\mathcal{N}_{\rm c}\approx 1$. Instead for the signal emitted during the PPI, that we focus on here, we have to consider a period of time of $\approx 20$ orbits, which implies $\mathcal N_{\rm c}$ is higher. As for the GW frequency, we expect this to be the Keplerian frequency, as in the case of TDEs, but multiplied by the PPI azimuthal number $m$. However, since for our system $m=1$, the expected frequency is just the Keplerian one.

\subsection{Amplitude of the maximum strain}
\label{sec:anal_study}
If we consider the disruption of a solar type star by a non-rotating $10^6$M$_{\sun}$ hole, we know from equation~\eqref{eq:beta} that $\beta$ varies in the range
\begin{align}
    1\leq\beta \textcolor{black}{\lesssim} 20.
\end{align}
Since, for a given star, the GW strain and frequency depend only on $\beta$, we have that these quantities range from $h \textcolor{black}{\approx} 10^{-22}$ and $f\textcolor{black}{\approx}10^{-4}\,\text{Hz}$, for $\beta=1$, to $h\textcolor{black}{\approx}2\times 10^{-21}$ and $f \textcolor{black}{\approx}9\times 10^{-3}\,\text{Hz}$, for $\beta=20$. From the GW strain it is possible to derive $h_{\rm c}$ through equation~\eqref{eq:strain_cycle}, where the number of cycles is given by (see equation~\ref{eq:numbercycle})
\begin{align}
   \mathcal{N}_{\rm c}\approx 57.
   \label{eq:n_c}
\end{align}
\begin{figure}
    \centering
    \includegraphics[width=\columnwidth, height=0.24\paperheight]{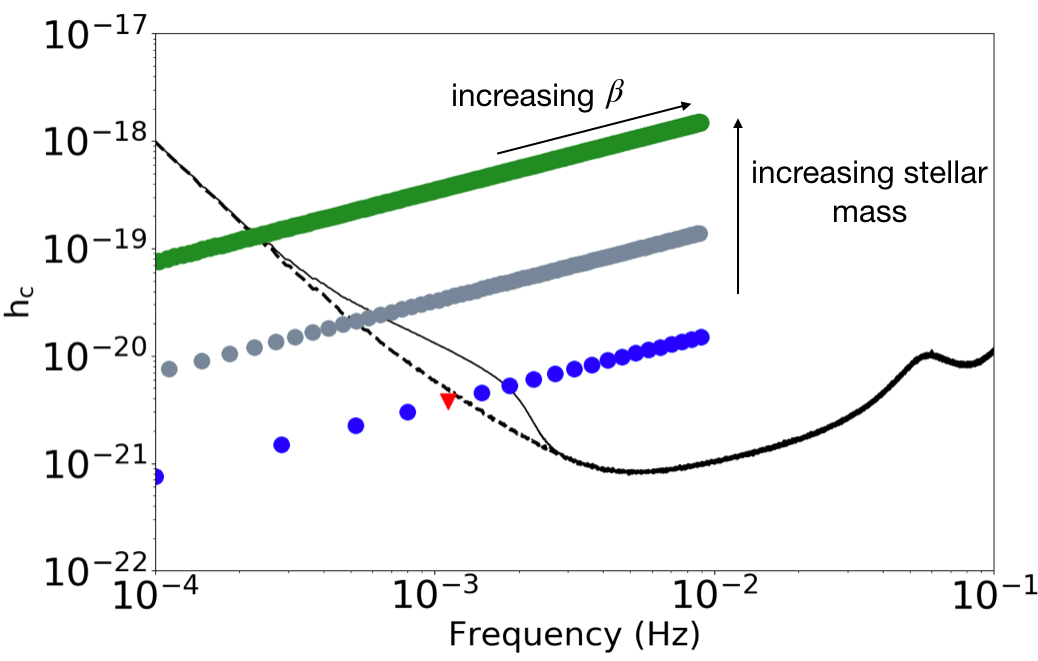}
    \caption{Analytical estimates of the GW characteristic strain
for a MS star disrupted by a non-rotating $10^6\text{M}_{\sun}$ hole, \textcolor{black}{with $\mathcal N_{\rm c}\approx 57$ (see equations~\ref{eq:strain_cycle} and paragraph \ref{sec:anal_study}), used to conjecture the characteristic strain of a TDE remnant unstable to the PPI for 20 orbits. This signal is} compared with respect to the sensitivity curve of LISA (black  curve\textcolor{black}{, with  the  solid one being the instrumental sensitivity curve + 4 year foreground, while the dashed one being  the instrumental noise}). We consider three different values of stellar mass, $M_*=1,10,100$M$_{\sun}$ in blue, grey and green respectively. \textcolor{black}{The red triangle represents the signal from a source which corresponds to the torus simulated by \citet{Nealon:18aa} (i.e. a torus resulting from a TDE with $\beta=5$)}. The penetration factor $\beta$ increases from left to right. The signal gets above the sensitivity curve when $\beta_{\rm cr}\approx 6,15,45$ for $M_*=1,10,100\text{M}_{\sun}$ respectively.}
    \label{fig:fig1}
\end{figure}
We obtain this number considering that the torus simulated by \citet{Nealon:18aa} shows the PPI at a frequency which is the Keplerian frequency of the star tidally disrupted (so calculated at $r_{\rm p}$), while the torus is located around $2r_{\rm p}$. So $\tau$ is 20 times the period at $2r_{\rm p}$. The peak signal expected from this source with respect to the LISA sensitivity curve is illustrated in figure \ref{fig:fig1}, where we have that the black curve is the sensitivity curve of the instrument, while the blue dots stand for the signal from a $1$M$_{\sun}$ \textcolor{black}{TDE remnant}. The parameter $\beta$ increases from left to right. We show with a red triangle the case simulated by \citet{Nealon:18aa}, which has a characteristic strain and a frequency given by
\begin{align}
    h_{\rm c}\approx 4\times 10^{-21},\,\,\,\, f=10^{-3}\,\text{Hz}.
\end{align}
The signal goes above the sensitivity curve of LISA \textcolor{black}{for $\beta \gtrsim 6$}. \textcolor{black}{To find this critical value of $\beta$, $\beta_{\rm cr}$, we infer from the plot the frequency at which the signal intersects the LISA sensitivity curve, and then we derive $\beta_{\rm cr}$ using equation ~\eqref{eq:freq_ms}, considering $m_*=1$.} Therefore, the case simulated by \citet{Nealon:18aa} appears not to be visible by the interferometer.

It is interesting to extend these calculations for a larger range of stellar masses. So, we investigate the signal from MS stars in the interval $1\leq m_*\leq 100$. However we have to consider, that, increasing $m_*$ also increases the minimum $\beta$ that is still in the frequency range visible to LISA\textcolor{black}{, since, as we can see from equation~\eqref{eq:freq_ms}, we have}
\begin{align}
    \textcolor{black}{\beta_{min}=\left(\frac{f_{\rm min}}{10^{-4}\,\text{Hz}}\right)^{2/3}m_{*}^{2/3}}.
\end{align}
\textcolor{black}{Also} in figure \ref{fig:fig1}, we have plotted the signal for \textcolor{black}{$m_*=10,100$ (grey and green respectively), where} $\beta$ increases from left to right. For $m_*=10$, we have considered $5\leq \beta \leq 92$, while for $m_*=100$ we have $22\leq\beta\leq 430$. \textcolor{black}{In particular, we have that the signal overtakes the instrument sensitivity curve for $\beta_{\rm cr}\approx 15$ if $m_*=10$, and $\beta_{\rm cr}\approx 45$ if $m_*=100$}.

It is interesting to note that for each mass the signal which corresponds to the highest value of $\beta$ is at a frequency of $\approx 9\times 10^{-3}\,\text{Hz}$, which is indeed approximately the frequency of the innermost stable orbit for a hole of $10^6M_{\sun}$. \textcolor{black}{This can be also seen from equation~\eqref{eq:freq_ms}, from which we derive that the maximum frequency, $f_{\rm max}$, is independent on $m_*$ and is given by} 
\begin{equation}
   \textcolor{black}{f\lesssim f_{\rm max}=f[\beta_{\rm max}] \simeq 9\,\text{mHz}\, M_6^{-1}},
\end{equation}
\textcolor{black}{where $\beta_{\rm max}=r_{\rm t}/2r_{\rm gh}$.}\\
\indent From these simple, order of magnitude estimates of the process, it seems that the TDE of a solar type star by a static $10^6$M$_{\sun}$ hole may be detectable by LISA, when we consider high values of $\beta$. However, we have to consider \textcolor{black}{three} important simplifications that probably make our calculations overestimate the real value:
\begin{enumerate}
    \item not all of the stellar mass is involved in the PPI, i.e. the amplitude of the PPI perturbations is $<1$;
    \item the torus is not located at $2r_{\rm p}$ but it is spreading out;
    \item \textcolor{black}{other effects, such as the MRI, would quench the PPI after a few orbits (see section \ref{sec:6}), reducing the number of cycles $\mathcal{N}_{\rm c}$}.
\end{enumerate}

\begin{figure*}
    \centering
    \includegraphics[width=0.46\textwidth]{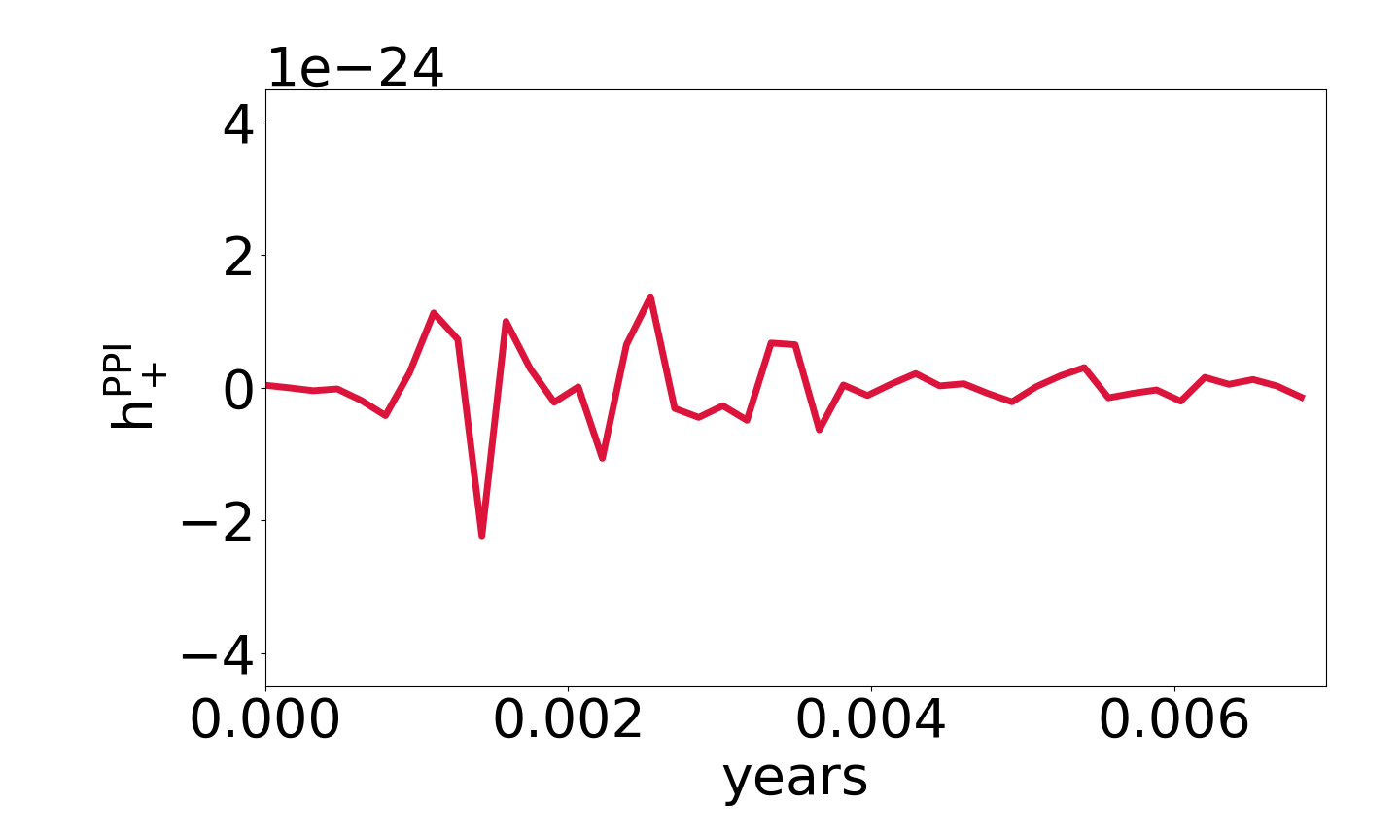}
\includegraphics[width=0.46\textwidth]{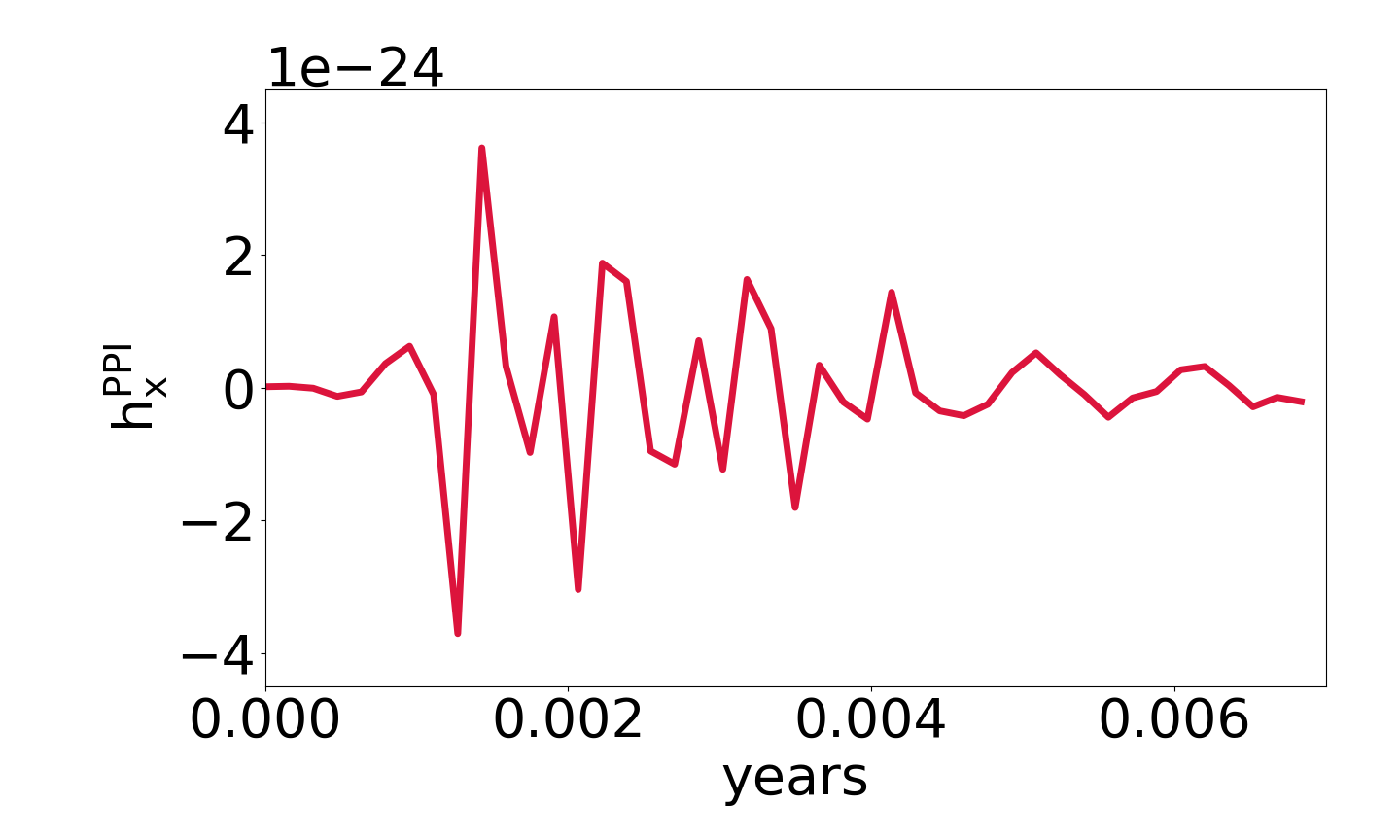}
\caption{GW waveforms plotted with respect to time $t$, expressed in years, along the $\pmb z$ direction of propagation. On the left panel we plot $\textcolor{black}{h^{\rm PPI}_{+}(t)}$, that reaches the peak value (in magnitude) of $\sim 2 \times 10^{-24}$ between $ 1\times 10^{-3}\,\text{yr}$ and $\sim 2\times 10^{-3}\,\text{yr}$. On the right panel we plot $\textcolor{black}{h^{\rm PPI}_\times(t)}$, that reaches the peak value of $\sim 4 \times 10^{-24}$ between $ 1\times 10^{-3}\,\text{yr}$ and $\sim 2\times 10^{-3}\,\text{yr}$. } 
    \label{fig:fig3}
\end{figure*}

Thus, it is reasonable to expect that our analytical analysis overestimates the GW strain associated with the torus. So it makes sense to assume that the strain associated with the PPI, $h^{\rm PPI}$, can be written as
\begin{align}
    h^{\rm PPI}=\xi  h,
\end{align}
where $\xi$ is a factor that varies in the range (0,1]. To quantify the effect of the assumptions listed above and find $\xi$ we have thus performed a numerical study described in Section \ref{sec:4}.

\section{Numerical study}
\label{sec:4}
We have repeated the simulation of \citet{Nealon:18aa}: using the \textsc{PHANTOM} code, we simulate a $1$M$_{\sun}$ torus of \mbox{$N=5\times10^6$} particles and we require the cross section parameter, $d$, to be
\begin{align}
    d=\frac{(r_+-r_-)}{2r_0}=1.15,
\end{align}
where $r_0=0.5\,\text{AU}\textcolor{black}{\approx} 2r_{\rm p}$, $r_{\rm p}$ being the pericenter of the \citet{Bonnerot:16aa}'s simulation. $r_0$ is the distance of the maximum density of the torus. First of all\textcolor{black}{,} we relax the particles in order to reduce numerical artefacts due to the fact that they are set on a grid, then we set the torus in a fixed Keplerian potential and we add $m=1$ density perturbation to have the PPI. As in \citet{Nealon:18aa} we also only implement viscosity in order to capture shocks. The simulation runs for $\sim 20$ orbits, and each period $T$ is given by 
\begin{align}
T=\frac{2\pi r_0^{3/2}}{\sqrt{GM_{\rm h}}}=3.5\times 10^{-4}\,\mbox{yr}.
\label{eq:tempo}
\end{align}
As in \citet{Nealon:18aa}, during the first 3 orbits an over-density develops in the torus that orbits with the Keplerian frequency. The PPI continues to grow until 5th-6th orbits, when the over-density reaches its peak and a shock has developed, which spreads from the inner to the outer radii. Then the overdensity remains beyond this and the shock decreases.\\
\indent Since we are interested in deriving $h_{+}$, $h_{\times}$ and $h$ associated with the system, we need to discretize equations~(\ref{eq:h+}) and ~(\ref{eq:hx}). We start by discretizing the momentum of inertia of the system, introduced with equation~\eqref{eq:mom_in}, as
\begin{align}
    M^{kl}=\int \,\text{d}^{3} x \rho x^{k}x^{l}=\int \,\text{d}m x^{k}x^{l} \Rightarrow M^{kl}=\sum_{\rm a}m_{\rm a}x_{\rm a}^{k}x_{\rm a}^{l},
    \label{eq:discr}
\end{align}
where $\rm a$ is the index that runs over the number of particles and $m_{\rm a}$ is the mass of the $\rm a$-th particle. Here we are only using the motion of the star material to calculate the GW signal (this is justified in appendix \ref{appendix:bh}). Since in the simulation we use same mass particles, we can write $m_{\rm a}=m$ and so we have
\begin{align}
  M^{kl}=m\sum_{\rm a}x_{\rm a}^{k}x_{\rm a}^{l}.
  \label{eq:disc_quad}
\end{align}
Then we estimate the derivative of equation~(\ref{eq:disc_quad}) numerically using central differencing
\begin{align}
\ddot M_{\rm j}^{kl}=\frac{M^{kl}_{\rm j+1}-2M^{kl}_{\rm j}+M^{kl}_{\rm j-1}}{\Delta t^2},
\end{align}
where $j$ is the index related to time. Finally we substitute $\ddot M_{11}$, $\ddot M_{22}$ and $\ddot M_{12}$ into equations~(\ref{eq:h+}) and (\ref{eq:hx}). 

\subsection{Numerical results}
\begin{figure}
    \centering
    \includegraphics[width=\columnwidth]{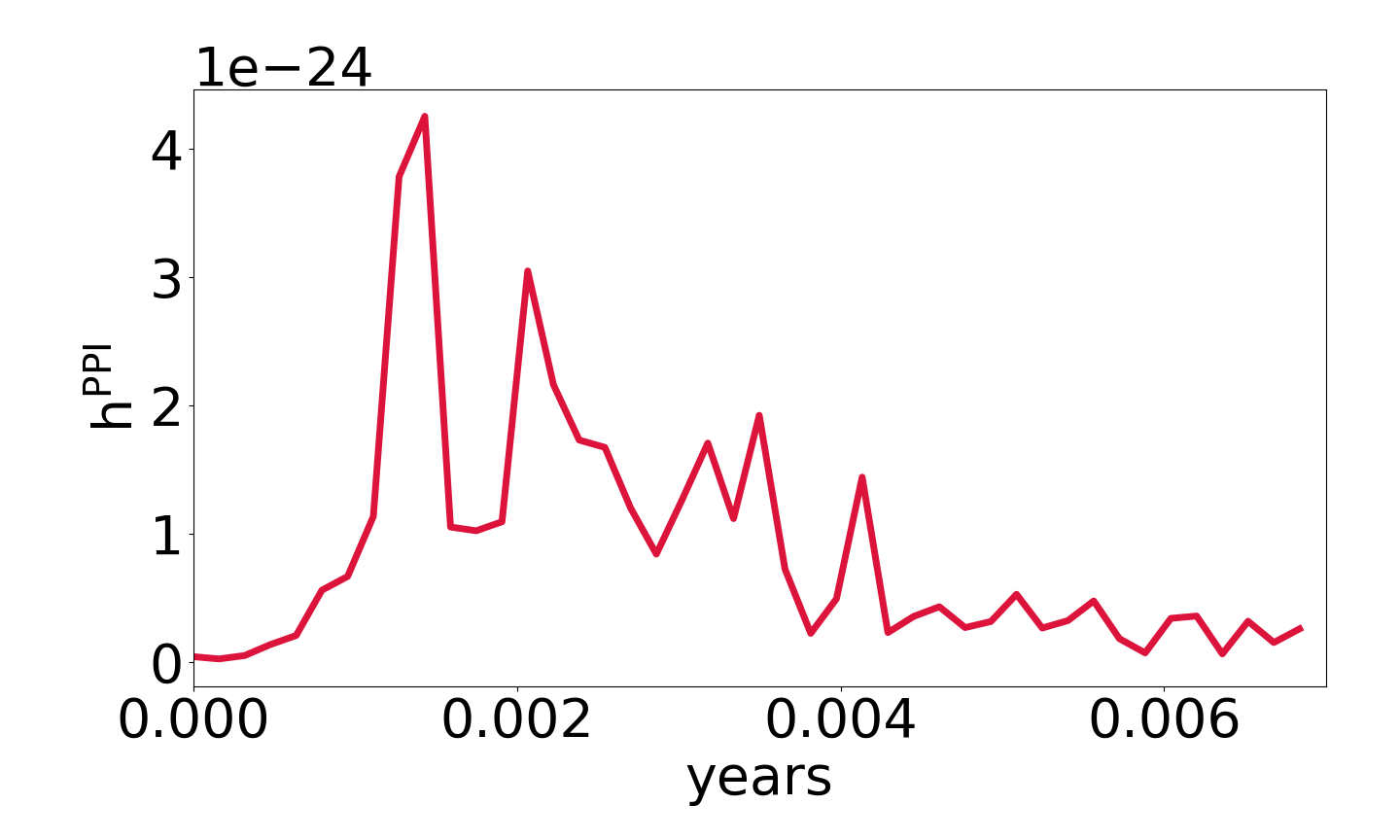}
    \caption{GW strain, $\textcolor{black}{h^{\rm PPI}(t)}$, plotted with respect to time $t$, expressed in years. The signal reaches the peak value of $ \sim 4 \times 10^{-24}$, two orders of magnitude lower than the analytical estimates of equation~\eqref{strain_ms}, between $\sim 1\times 10^{-3}\,\text{yr}$ and $\sim 2\times 10^{-3}\,\text{yr}$.}
    \label{fig:fig4}
\end{figure}

We numerically derive $\textcolor{black}{h^{\rm PPI}_+}$, $\textcolor{black}{h^{\rm PPI}_\times}$ from the SPH simulations for different directions, \textcolor{black}{where the superscript `PPI' stands for the strain measured from the simulation}. In particular in figure \ref{fig:fig3} we show $\textcolor{black}{h^{\rm PPI}_+}$ (left) and $\textcolor{black}{h^{\rm PPI}_\times}$ (right), calculated for the wave propagating in the $\pmb z$ direction, that is the direction perpendicular to the stellar orbit. Both waveforms reach the peak between $\sim 1\times 10^{-3}\,\text{yr}$ and $\sim 2\times 10^{-3}\,\text{yr}$, that is around the fifth and sixth orbit, when the overdensity is stronger (see \citealt{Nealon:18aa}), and in particular the peak value for the $+$ polarization is lower than the one of the $\times$ polarization by a factor $\sim 2$.\\
\indent In figure \ref{fig:fig4} we plot the strain of the wave $\textcolor{black}{h^{\rm PPI}}$ (see equation~\ref{eq:strain}) with respect to the time $t$ in years, along the direction of propagation $\pmb z$. It is interesting to note that the peak value is $\sim 4 \times 10^{-24}$, which is two orders of magnitude lower than the expected analytical estimate of $\sim 5 \times 10^{-22}$ for the raw strain, shown in equation~\eqref{strain_ms}. So for the torus formed after a TDE of a solar mass star, disrupted by a static $10^6\text{M}_{\sun}$ hole, with $\beta=5$, we have that $\xi\sim 10^{-2}$ and so
\begin{align}
    \textcolor{black}{h^{\rm PPI}}=10^{-2}h.
\end{align}
If we assume that the same scaling factor holds also for different $\beta$ and for different stellar masses, we can extrapolate our results by shifting down by $10^{-2}$ the signals shown in figure \ref{fig:fig1}. Thus, the expected signal would be visible only for $M_* \gtrapprox 10$M$_{\sun}$ \textcolor{black}{and for values of $\beta$ higher than $\beta_{\rm cr}$ defined above}.

\subsection{Resolution test}
\begin{figure}
    \centering
    \includegraphics[width=\columnwidth]{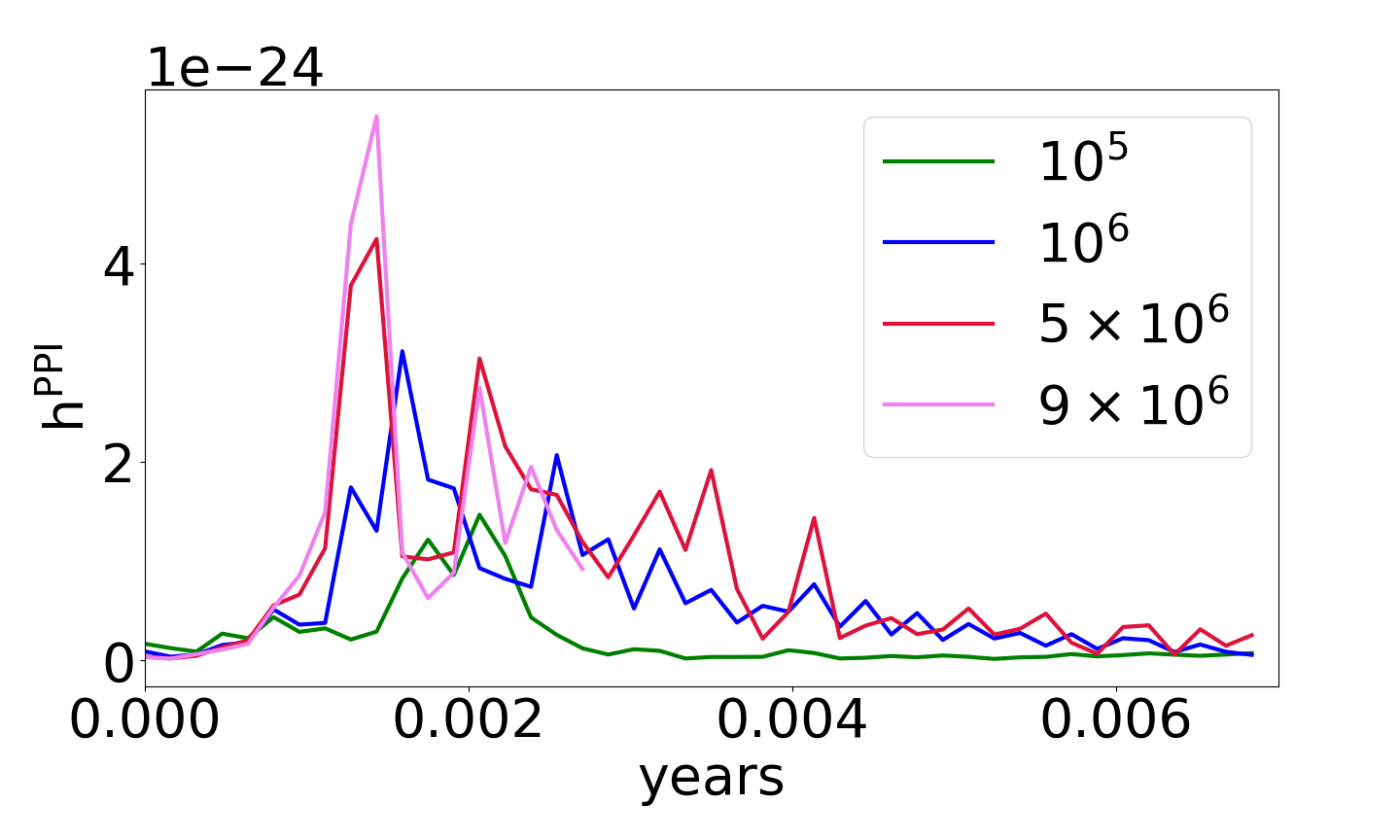}
    \caption{Strain $\textcolor{black}{h^{\rm PPI}}$ plotted with respect to the time for different resolution\textcolor{black}{s}. From bottom to top: $10^5$ (green), $10^6$ (blue), $5\times 10^6$ (red) and $9\times 10^6$ (pink) particles. It is possible to see the convergence from $5\times 10^6$ particles.}
    \label{fig:fig5}
\end{figure}

We have run the same simulation with different resolutions. In particular we have \textcolor{black}{used} $10^5$, $10^6$, $5\times 10^6$ and $9\times 10^6$ particles. For each simulation we plot the strain of the wave with respect to the time, as shown in figure~\ref{fig:fig5}. From this plot, it seems that, already with $5\times 10^6$ particles, there is convergence. In particular the peak of the strain is in similar position for the two highest resolutions. The curve corresponding to $9\times 10^6$ particles has only been run for long enough to demonstrate convergence. We thus regard our estimates from the simulation to be robust.

\section{Discussion}
\label{sec:5}
According to our analytical study, the maximum peak value associated to 1M$_{\sun}$ remnant disrupted by a static $10^6$M$_{\sun}$ hole, with $\beta=5$, is just under the sensitivity curve of LISA. However, if we increase $\beta$, the signal becomes visible to the interferometer ($\beta \gtrsim 6$). Similar results can be obtained also for all the stars in the range $1\leq m_*\leq 100$, but, while increasing $m_*$, we have to check that the GW frequency remains in the range accessible to LISA.\\
\indent The estimates made for the raw strain are similar to the ones in \citet{Kobayashi_2004}. However we need to consider that $h_{\rm c}$ is increased by a factor $\mathcal{N}_{\rm c}^{1/2}\approx 7.5$. In fact, since the GW signal of a TDE is a burst signal, we have that $\mathcal{N}_{\rm c}\approx 1$, which means $h_{\rm c}\approx h$. On the contrary, for the GW signal associated to the PPI in thick accretion discs, which is emitted on a longer time interval (in our system, 20 orbits), we have $h_{\rm c}\approx 7.5 h$. \\
\indent As for the numerical study, we show that the GW \textcolor{black}{strain} derived numerically from a 3D SPH simulation, performed in a classical frame, is two orders of magnitude lower than the analytical expectation
\begin{align}
    h^{\rm PPI}=10^{-2}h.
\end{align}
\indent This difference is due to the assumptions that we have made for our analytical analysis. First of all, the derivation of the strain is made for a tidally disrupted star, which is approximated by a point mass particle. Instead, the real system is made of a torus, that is spreading out due to the PPI. For this reason, considering the disc as a point where all the mass is located is not precise enough. Moreover, with our analytical study, we assume that all the total gas mass of the disc, since it is enclosed in the particle, undergoes the PPI, but in the real physical system in general only a part of the mass is involved. This is another feature that we are neglecting in the first part of our study. \\
\indent It is interesting to compare the results we obtain numerically with the ones of \citet{Kiuchi:11aa}. \textcolor{black}{In contrast to our work, t}hey show that their numerical results are one order of magnitude lower than the analytical estimates (figure 4 of \citealt{Kiuchi:11aa}), not two like in our case. However their study present some differences with respect to ours \textcolor{black}{that may justify this discrepancy}. First of all, they consider a disc much more massive, formed in a completely different scenario. In particular, the torus-SMBH mass ratio is bigger than in our case, which implies that it is reasonable to expect a significant contribution to the GW signal also from the BH. Then, they consider a self-gravitating torus. Self gravity is not a requirement of the PPI, since this instability can occur in any torus that is not accreting. However, self-gravity \textcolor{black}{plays} an important role in the formation of clumps of matter inside the torus, that are the sources of the GWs.  Moreover, we also need to consider that for sources with a non negligible self-gravity, it is not clear that the derivation made in paragraph \ref{ssec:gws} still holds (see \citealt{Buonanno:07aa}). For detailed discussions about the role of self-gravity in relativistic discs see \citet{Korobkin:11aa} and \citet{Mewes:16aa}. Finally they use a grid code with GR. \citet{Kobayashi_2004} have shown that for TDEs with low $\beta$, GR does not change the estimates of GW emission in a significant way. However, this could be different in the case of accretion discs, which are very massive (as in \citealt{Kiuchi:11aa}) and for this reason have a non-negligible interaction with the SMBH.

\section{Conclusions}
\label{sec:6}
In this work we have studied the GW signal associated with the PPI in a thick disc, with a shallow specific angular momentum profile and an inner and outer radii well defined. This disc has just resulted from the TDE of a $1$M$_{\odot}$ star around a non rotating SMBH of $10^6 $M$_{\odot}$, with $\beta=5$.\\
\indent First of all we have made some analytical estimates of the maximum amplitude, to see if the signal could be above the LISA sensitivity curve. From this study we have found that the signal might be visible for discs with masses in the range $[1$M$_{\sun},100$M$_{\sun}]$, for high values of  $\beta$, \textcolor{black}{in particular $\beta \gtrsim 6 \,\text{for}\, 1\text{M}_{\sun}$, $\beta \gtrsim 15 \,\text{for}\, 10\text{M}_{\sun}$, and $\beta \gtrsim 45 \,\text{for}\, 100\text{M}_{\sun}$}.\\
\indent Then we have performed a numerical 3D SPH simulation, using the \textcolor{black}{\textsc{PHANTOM}} code, and compared the numerical results with the previous estimates. We have found that the numerical maximum strain is two orders of magnitude lower than the analytical one. This lowering can be justified considering that, during our analytical analysis, we neglect the disc is spreading out due to the PPI, and that not all the disc mass is involved in the instability. \\
\indent \textcolor{black}{It is important to remember that in our simulation we do not consider magnetic fields. The inclusion of magnetic fields from the beginning could suppress the GW emission or even avoid it, according to the particular structure of the disc after circularization. In this regard, \citet{Bugli:18aa} have illustrated that the MRI can allow an initial formation of the PPI, but the disc needs to already show a dominant $m=1$ mode. The final $m=1$ mode present in \citet{Bonnerot:16aa}, that justifies the perturbation in our model as in \citet{Nealon:18aa}, may not occur if magnetic fields are included \textit{a priori}.}\\
\indent In conclusion, our numerical study suggests that this source could still be detectable by LISA, if we consider tori with masses in the range $[10$M$_{\sun},100$M$_{\sun}]$, resulting after \textcolor{black}{deeply penetrating} TDEs.

\textcolor{black}{\section*{Acknowledgements}
MT and GL have received funding from the European Union's Horizon 2020 research and innovation programme under the Marie Sk\l{}odowska-Curie grant agreement NO 823823 (RISE DUSTBUSTERS project). RN has received funding from the European Research Council (ERC) under the European Union's Horizon 2020 research and innovation programme (grant agreement No 681601)}.

\bibliographystyle{mnras}




\appendix
\section{The contribution of the black hole to the gravitational wave signal}
\label{appendix:bh}

In equation~\eqref{eq:discr}, we calculate the GW signal of the system only taking into account the contribution of the torus and not of the hole. This can be explained in the following ways.\\
\indent First of all we can approximate the system SMBH-torus as a binary system where we have that the total mass of the system, $M=M_{\rm h}+M_{\rm d}$, where $M_{\rm d}$ is the mass of the disc, can be approximated with the mass of the black hole, since we have that $M_{\rm h}$ is between $10^6$ and $10^{4}$ times the mass of the torus, if we consider tori with masses in the range $[1\text{M}_{\sun},100\text{M}_{\sun}]$. The reduced mass of the system, defined as
\begin{align}
\mu=\frac{M_{\rm h}M_{\rm d}}{M_{\rm h}+M_{\rm d}},
\end{align}
is approximately equal to the mass of the disc, for the same reasons above. So it is reasonable to say that the position of the centre of mass of the system corresponds to the position of the black hole, since the black hole is much more massive than the torus. These considerations suggest that the mass of the black hole is not a significant factor in the amplitude of the GWs, and as such we do not consider it in our derivation of equation~\eqref{eq:discr}.\\
\indent A more formal way to show that the SMBH does not contribute to the GW signal emission in our study, is the the following. In the centre of mass frame, we have that
\begin{align}
    M_{\rm h}r_{1}\approx M_{\rm d}r_{2},
\end{align}
where $r_{1}$ and $r_{2}$ are the displacements from the centre of mass of the hole and the disc, respectively. If we want to derive the moment of inertia of the black hole we have
\begin{align}
    M_{\rm h}r_{1}^2=M_{\rm d}r_{2}^2\left(\frac{M_{\rm d}}{M_{\rm h}}\right).
\end{align}
In particular, if we assume $M_{\rm h}=10^6\text{M}_{\sun}$ and $M_{\rm d}= 1\text{M}_{\sun}$, we have that the moment of inertia of the hole becomes
\begin{align}
 M_{\rm h}r_{1}=10^{-6}M_{\rm d}r_{2}^2,
\end{align}
while if $M_{\rm d}=100$M$_{\sun}$ we have 
\begin{align}
 M_{\rm h}r_{1}=10^{-4}M_{\rm d}r_{2}^2.
\end{align}
The above illustrates that the GW signal emitted by the SMBH is negligible with respect to that emitted by the material in the disc. Hence when we calculate the GW signal, we only consider the mass in the disc.

\bsp	
\label{lastpage}
\end{document}